\documentclass[preprint]{aastex}
\newcommand{\kms}{km~s$^{-1}$}
\usepackage{natbib}
\usepackage{graphicx}
\usepackage{amssymb}
\usepackage{txfonts}
\begin{document}

\title{Water Vapor in the Inner 25 AU of a Young Disk around a Low-Mass Protostar\footnote{Based on observations carried out with the IRAM Plateau de Bure Interferometer. IRAM is supported by INSU/CNRS (France), MPG (Germany) and IGN (Spain).}}
\author{Jes K. J{\o}rgensen}
\affil{Argelander-Institut f\"{u}r Astronomie, University of Bonn, Auf dem H\"{u}gel 71, D-53121 Bonn, Germany\footnote{\emph{Current address:} Centre for Star and Planet Formation, Natural History Museum of Denmark, University of Copenhagen, {\O}ster Voldgade 5-7, DK-1350 Copenhagen K, Denmark}}
\email{jes@snm.ku.dk}
\and
\author{Ewine F. van Dishoeck}
\affil{Leiden Observatory, Leiden University, PO Box 9513, NL-2300 RA
Leiden, The Netherlands \\ \& Max-Planck Institut f\"ur extraterrestrische
Physik, Giessenbachstrasse, D-85748 Garching, Germany}
\email{ewine@strw.leidenuniv.nl}

\begin{abstract}
  Water is one of the key molecules in the physical and chemical
  evolution of star- and planet-forming regions.  We here report the
  first spatially resolved observation of thermal emission of (an
  isotopologue of) water with the Plateau de Bure Interferometer
  toward the deeply embedded Class 0 protostar NGC~1333-IRAS4B. The
  observations of the H$_2^{18}$O $3_{1,3}-2_{2,0}$ transition at
  203.4~GHz resolve the emission of water toward this source with an
  extent of about 0.2$''$ corresponding to the inner 25~AU
  (radius). The H$_2^{18}$O emission reveals a tentative velocity
  gradient perpendicular to the extent of the protostellar outflow/jet
  probed by observations of CO rotational transitions and water
  masers. The line is narrow $\approx$~1~\kms\ (FWHM), significantly
  less than what would be expected for emission from an infalling
  envelope or accretion shock, but consistent with emission from a
  disk seen at a low inclination angle. The water column density
  inferred from these data suggests that the water emitting gas is a
  thin warm layer containing about 25 $M_{\rm Earth}$ of material,
  0.03\% of the total disk mass traced by continuum observations.
\end{abstract}


\maketitle

\section{Introduction}\label{introduction}
Water is one of the most important molecules in star-forming regions:
it is a dominant form of oxygen, is important in the energy balance,
and is \emph{ultimately} associated with the formation of planets and
emergence of life. Thus, following the water ``trail'' from collapsing
clouds to protoplanetary disks is a fundamental problem in astronomy
and astrochemistry.  In the cold and quiescent regions the gaseous
water abundance is low, only $10^{-9}-10^{-8}$
\citep[e.g.,][]{bergin02h2o}, but in regions with intense heating or
active shocks, its abundance can reach 10$^{-4}$ with respect to H$_2$
-- comparable to or higher than that of CO
\citep[e.g.,][]{harwit98}. Which mechanism is most important for
regulating the H$_2$O abundance in low-mass protostars is still
heavily debated, however. Is it passive heating of the collapsing
envelope by the accretion luminosity from forming stars
\citep[e.g.,][]{ceccarelli98h2o,maret02}, or shocks caused either by
protostellar outflows \citep[e.g.,][]{nisini99} or related to ongoing
accretion onto circumstellar disks \citep{watson07}? H$_2$O is also a
key molecule in the chemistry in regions of star formation: in large
parts of the cold and dense envelopes around low-mass protostars,
H$_2$O is the dominant constituent of the icy mantles of dust grains
\citep[e.g.,][]{whittet88,boogert08}. Its evaporation at temperatures
higher than 90--100~K determines at what point water itself and any
complex organic molecules, formed in these ice mantles, are injected
into the gas-phase.

This discussion has received new fuel with the detection of
surprisingly strong highly excited H$_2$O lines at mid-infrared
wavelengths with the {\it Spitzer Space Telescope}'s infrared spectrograph,
IRS, toward one deeply embedded Class 0 protostar, NGC~1333-IRAS4B, by
\cite{watson07}. Based on the high critical density of the observed
lines and temperature ($\sim 170$~K), \citeauthor{watson07} argue that
the water emission observed toward this source has its origin in an
accretion shock in its circumstellar disk. Those data could not
spatially or spectrally resolve the water emission, however.

In this letter we present observations at 203 GHz of the H$_2^{18}$O
isotopologue at high-angular resolution (0.5$''$) using the Institut
de Radioastronomie Millim\'etrique (IRAM) Plateau de Bure
Interferometer to determine the origin of water emission in low-mass
protostars.  This isotopic line is a useful tracer of H$_2$O as it can
be detected and imaged at high angular resolution from the ground
under good weather conditions
\citep[e.g.,][]{jacq88,gensheimer96,vandertak06}. Its upper level
energy of 203.6 K is well matched to the observed excitation
temperature of water seen by {\it Spitzer}.

\section{Observations}
We observed NGC~1333-IRAS4B (IRAS4B in the following;
$\alpha$=03$^{\mathrm h}$29$^{\mathrm m}$12\fs00, $\delta$=+31$^\circ$13\arcmin08\farcs1 [J2000]
\citealt{prosacpaper})\footnote{In this paper we adopt a distance of
  250~pc for NGC~1333-IRAS4B (see discussion in \cite{enoch06}).}
using the 6 element Institut de Radioastronomie Millim\'etrique (IRAM)
Plateau de Bure Interferometer. The receivers were tuned to the
para-H$_2^{18}$O $3_{1,3}-2_{2,0}$ transition at
203.407498~GHz (1.47~mm). The correlator was set up with one unit with
a bandwidth of 36~MHz (53~\kms) centered on this frequency providing a
spectral resolution on 460 channels of 0.078~MHz (0.11~\kms). The
source was observed in two configurations: in the C configuration on
02 December 2008 and in the B configuration on 11 and 13 January
2009. About 11 hours were spent in each configuration (including time
used on gain calibrators etc.). When combined these two configurations
cover baselines with lengths from 17.8 to 452~m (12 to
308~k$\lambda$).

The data were calibrated and imaged using the CLIC and MAPPING
packages from the IRAM GILDAS software. The calibration followed the
standard approach: the absolute flux calibration was established
through observations of MWC~349, the bandpass by observations of the
strong quasar 3c454.3 and the complex gains by regular observations of
the nearby quasar J0336+323 (approximately 0.8~Jy at
1.45~mm). Integrations with clearly deviating amplitudes and/or phases
were flagged and the continuum was subtracted prior to Fourier
transformation of the line data. With natural weighting the resulting
beam size is 0.67\arcsec $\times$ 0.55\arcsec at a position angle of
36.7$^\circ$; the field of view is 25\arcsec\ (HPBW) at 1.45~mm. The
resulting RMS noise level is 11~mJy~beam$^{-1}$~channel$^{-1}$ for the
line data using natural weighting. The continuum sensitivity is
limited by the dynamical range of the interferometer and the resulting
RMS noise level is a few~mJy~beam$^{-1}$.

\section{Results}
\begin{figure}
\resizebox{\hsize}{!}{\includegraphics{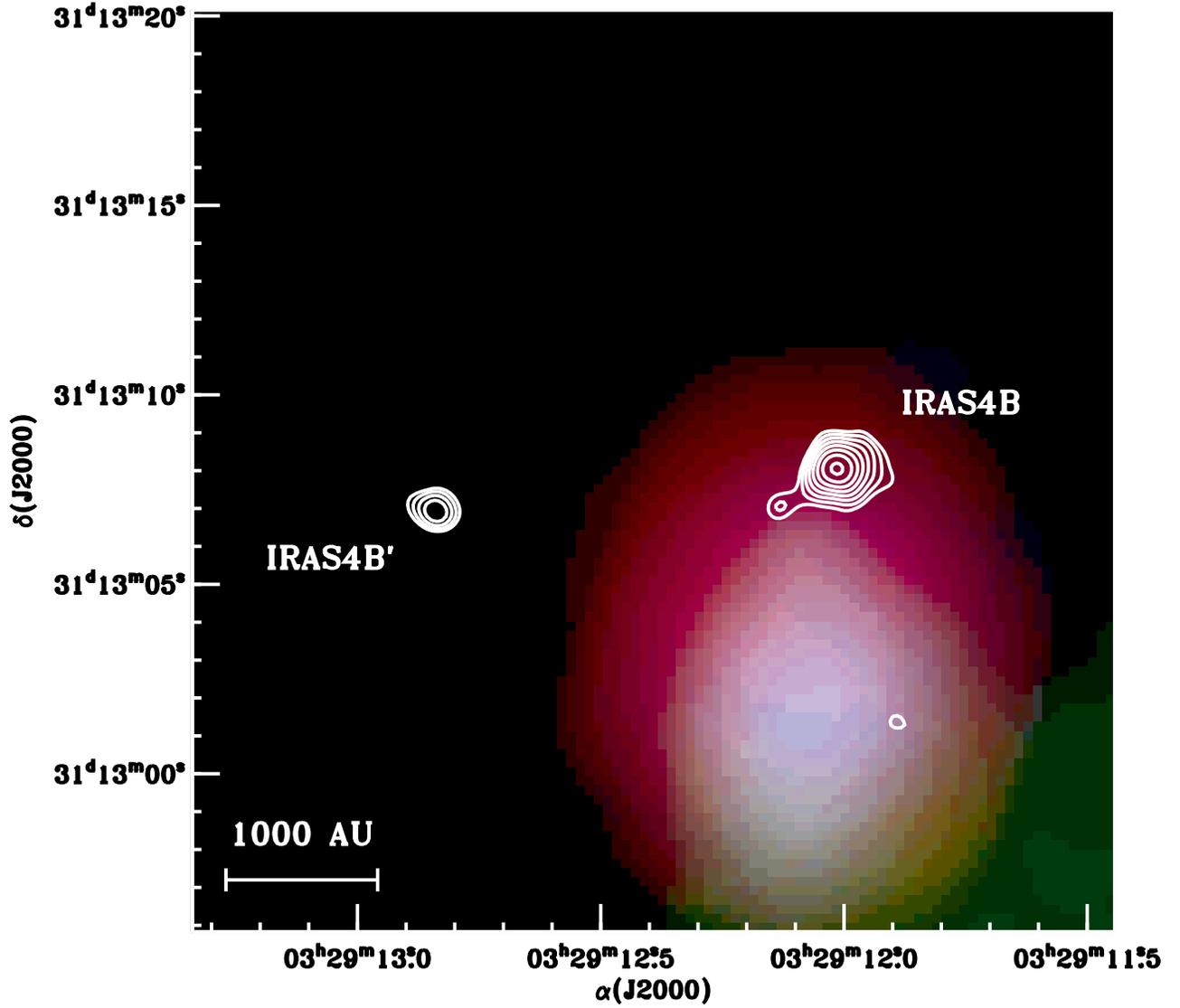}}
\caption{Continuum image of IRAS4B and IRAS4B$'$ from the IRAM PdBI
  observations (contours shown in logarithmic steps from 0.015 to
  0.25~Jy~beam$^{-1}$ overlaid on Spitzer Space Telescope mid-infrared
  image \citep{perspitz,gutermuth08ngc1333} with 4.5~$\mu$m emission shown in blue,
  8.0~$\mu$m in green and 24~$\mu$m in red.}\label{iras4b_cont}
\end{figure}

Fig.~\ref{iras4b_cont} shows the continuum image of the observed
region around IRAS4B. As seen both IRAS4B and its nearby companion
IRAS4B$'$ are clearly detected in the image. Table~\ref{cont_table}
lists the results of elliptical Gaussian fits to the two sources: both
are resolved with fluxes in agreement with the results from
\cite{prosacpaper} assuming that it has its origin in thermal dust
continuum emission with $F_\nu\propto \nu^{\alpha}$ with
$\alpha\approx 2.5-3$. The continuum peak is clearly offset by
5--7$''$ from the emission at 3.6--24~$\mu$m seen in the Spitzer Space
Telescope images of IRAS4B; an indication that the Spitzer emission
has its origin in material heated by the protostellar outflow even at
24~$\mu$m (see also \citealt{scubaspitz} and Fig.~2 of
\citealt{allenppv}).

\begin{table}\centering
  \caption{Parameters for IRAS4B and IRAS4B$'$ from
    elliptical Gaussian fits to their continuum emission.}\label{cont_table}
\begin{tabular}{lll}\hline\hline
                              & IRAS4B                   & IRAS4B$'$ \\ \hline
Flux                          &                0.586~Jy     &      0.128~Jy     \\
R.A. (J2000)                  &          03:29:12.01     &  03:29:12.84   \\
DEC (J2000)                          &          31:13:08.07     &  31:13:06.93   \\
Extent$^{a}$     &      0.80\arcsec$\times$0.54\arcsec\ ($-65^\circ$)   &      0.56\arcsec$\times$0.45\arcsec\ ($-86^\circ$) \\ \hline
\end{tabular}

$^{a}$Size of Gaussian from fit in $(u,v)$-plane (i.e., deconvolved FWHM size) and position angle of major axes (in parentheses).
\end{table}

\begin{figure}
\resizebox{0.5\hsize}{!}{\includegraphics{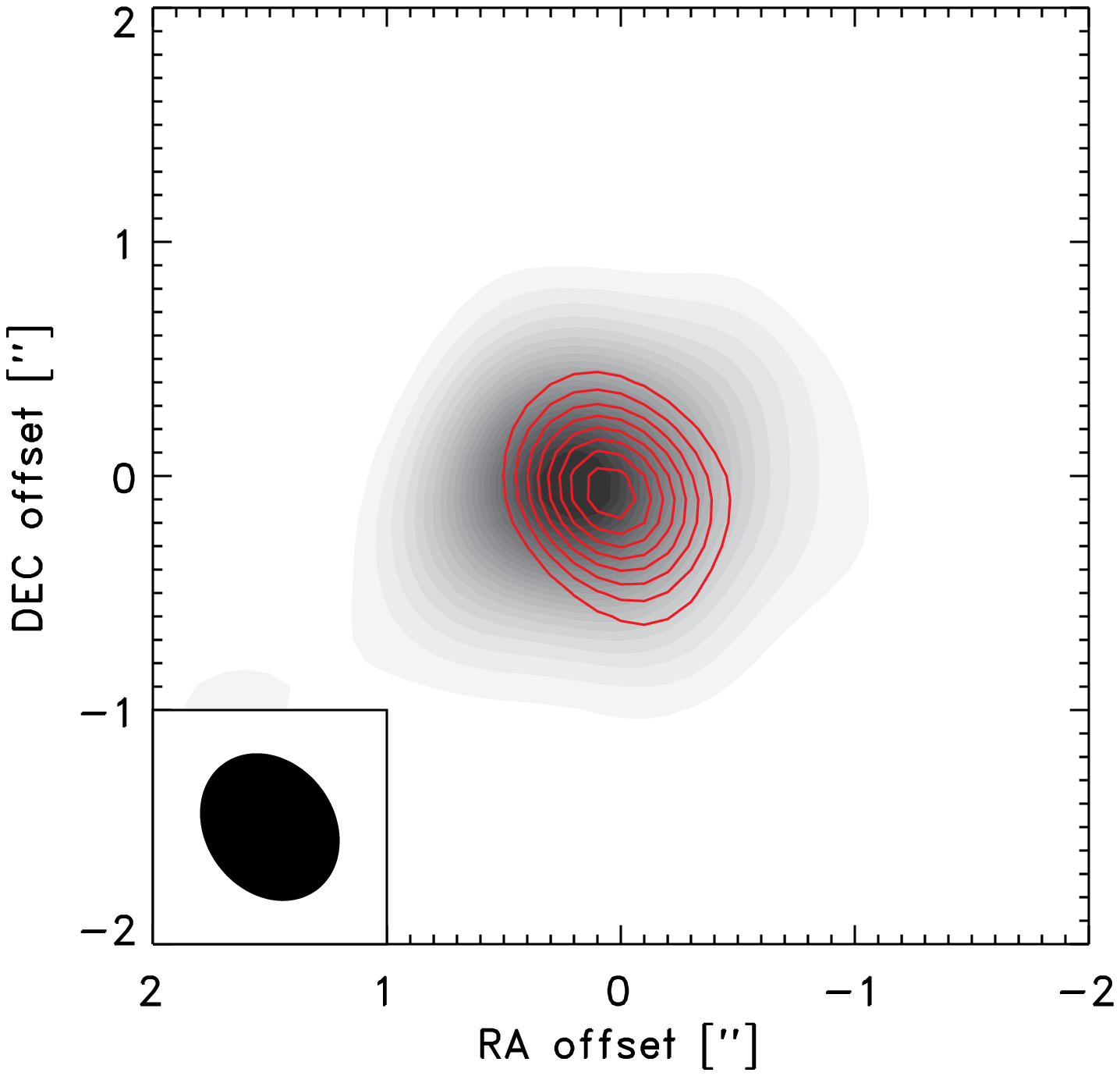}}
\resizebox{0.5\hsize}{!}{\includegraphics{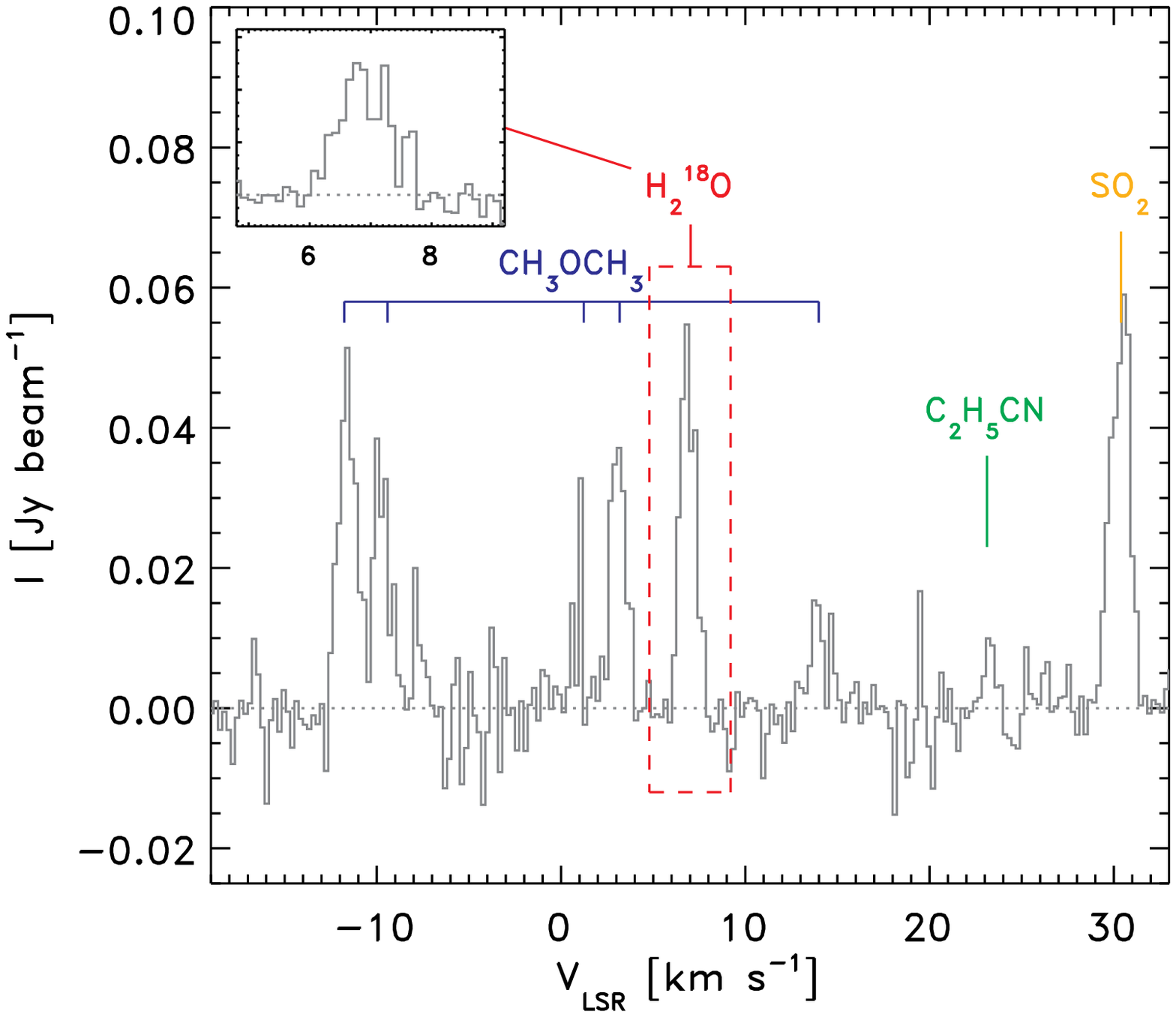}}
\caption{\emph{Left:} Integrated emission of the H$_2^{18}$O line
  (contours in steps of 5~mJy~beam$^{-1}$~\kms\ starting at
  10~mJy~beam$^{-1}$~\kms) plotted over the continuum emission
  (grey-scale). \emph{Right:} Spectrum extracted in the central
  $0.6''\times 0.5''$ beam toward the continuum position for
  IRAS4B. The detected lines are indicated at the position of their
  catalog rest frequency corrected for the 7.0~\kms\ systemic velocity
  of IRAS4B. The spectrum has been binned to twice the observed
  resolution The insert shows a blow-up of the H$_2^{18}$O line at the
  original resolution.}\label{spectrum}
\end{figure}

Fig.~\ref{spectrum} shows the spectrum toward the continuum peak of
IRAS4B. A number of lines are clearly detected as listed in
Table~\ref{line_id} -- including the targeted H$_2^{18}$O
$3_{1,3}-2_{2,0}$ line. For the line identification we used the JPL
\citep{pickett98} and CDMS \citep{mueller01,mueller05} databases and
cross-checked those with the online Splatalogue compilation. Since all
of the lines are narrow little line-blending occurs, in contrast with
high-mass star-forming regions. Also, all the assigned lines are
centered within 0.1--0.2~\kms\ of the systemic velocity of 7.0~\kms\
of IRAS4B. Most prominent in the spectrum are lines of dimethyl ether,
CH$_3$OCH$_3$, with 5 identified transitions. In addition, lines of
sulfur dioxide, SO$_2$, and water, H$_2^{18}$O, are clear detections
with a fainter line of ethyl cyanide, C$_2$H$_5$CN, also
present. Concerning the confidence of the assignment of the
H$_2^{18}$O line: according to the Splatalogue compilation no other
transitions fall within $\pm$1~\kms\ of the location of the
H$_2^{18}$O line. Offset by 1--2~\kms\ are transitions of
$^{13}$CH$_2$CHCN, $^{13}$CH$_3$CH$_2$CN and (CH$_3$)$_2$CO: the two
former can be ruled out because of the lack of additional components
which should have been observable at larger velocity offsets, whereas
the latter has a very low intrinsic line strength.

\begin{deluxetable}{llllllll}
\tablecaption{Identified lines and results from Gaussian fit to emission in $(u,v)$ plane.}\label{line_id}
\tablehead{\colhead{Molecule}      & \colhead{Transition}           & \colhead{Frequency} & \colhead{$E_u$}  & \colhead{$\mu^2S$}  & \colhead{Flux\tablenotemark{a}}  & \colhead{Offset ($\Delta\alpha,\Delta\delta$)\tablenotemark{b}}           & \colhead{Size\tablenotemark{c}} \\
\colhead{}      & \colhead{}           & \colhead{[GHz]} & \colhead{[K]}  & \colhead{}  & \colhead{[Jy~\kms]}         & \colhead{[\arcsec]}           & \colhead{[\arcsec]}}
\startdata
H$_2^{18}$O   & $3_{1,3}-2_{2,0}$    & 203.407498      & 203.7  &  0.344    & 0.081  & ($-$0.11; $-$0.02) & 0.20\arcsec \\
CH$_3$OCH$_3$ & $3_{3,1}-2_{2,1}$ EA & 203.402779      &  18.12 &  9.29     & 0.035  & ($-$0.14; $-$0.07) & 0.48\arcsec \\
              & $3_{3,1}-2_{2,1}$ EE & 203.410112      &  18.12 & 31.1      & 0.088  & ($-$0.14; $-$0.02) & 0.46\arcsec \\
              & $3_{3,0}-2_{2,1}$ AE & 203.411431      &  18.12 & 20.0      & 0.044  & ($-$0.12; $-$0.02) & 0.31\arcsec \\
              & $3_{3,0}-2_{2,1}$ AA & 203.418656      &  18.12 & 33.4      & 0.12   & ($-$0.12; $+$0.03) & 0.67\arcsec \\
              & $3_{3,0}-2_{2,1}$ EE & 203.420253      &  18.12 & 22.3      & 0.094  & ($-$0.16; $+$0.00) & 0.25\arcsec \\
C$_2$H$_5$CN  & $23_{2,22}-22_{2,21}$ & 203.396581      & 122.3  & 338       & 0.045  & ($+$0.02; $+$0.00) & 0.37\arcsec \\
SO$_2$        & $12_{0,12}-11_{1,11}$ & 203.391550\tablenotemark{d}&  70.12 & 22.5      & 0.12   & ($-$0.11; $-$0.02) & 0.41\arcsec \\ \hline
\enddata

\tablenotetext{a}{Total flux from fitting circular Gaussian to integrated line emission in the $(u,v)$ plane. For conversion to brightness temperatures, the gain of the interferometric observations with the current beam size is 0.0126~Jy~K$^{-1}$ -- i.e., implying an integrated line strength for the H$_2^{18}$O line of 6.5~K~\kms.}
\tablenotetext{b}{Peak offset with respect to continuum peak estimated from Gaussian fit.}
\tablenotetext{c}{Extent of emission (FWHM) from Gaussian fit.}
\tablenotetext{d}{Catalog frequency uncertain (accuracy $\pm 0.1$MHz). The observed SO$_2$ peak indicates a 
frequency higher by 0.1~MHz than the tabulated one.}
\end{deluxetable}

For each of the detected lines we fit a circular Gaussian in the
$(u,v)$ plane to the line emission integrated over the widths of the
lines to zero intensity (about $\pm 1$~\kms). The derived fluxes, peak
positions and sizes are given in Table~\ref{line_id} as well. All the
detected lines are marginally resolved with deconvolved sizes of
0.2$''$--0.6$''$, corresponding to radii of only 25 to
75~AU. Interestingly, most of the detected lines are offset by about
0.10--0.15\arcsec from the continuum position in right ascension --
with the exception of C$_2$H$_5$CN. The peaks of the molecular
emission still fall within the deconvolved extent of the continuum
emission, though. The images of the complex organic molecules and
SO$_2$ will be discussed elsewhere.

\section{Discussion}
\subsection{Velocity field}
\begin{figure}
\resizebox{0.8\hsize}{!}{\includegraphics{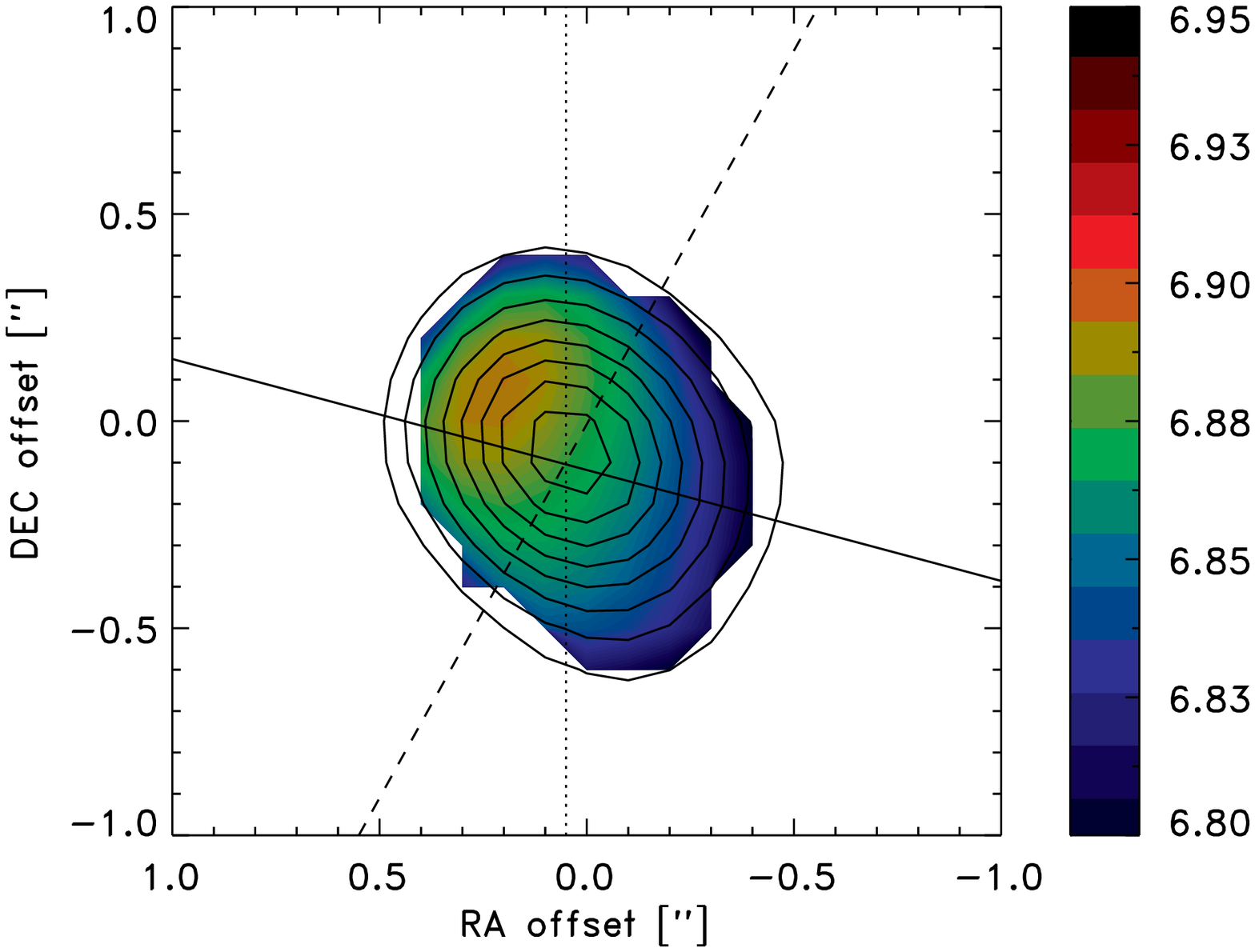}}
\resizebox{0.8\hsize}{!}{\includegraphics{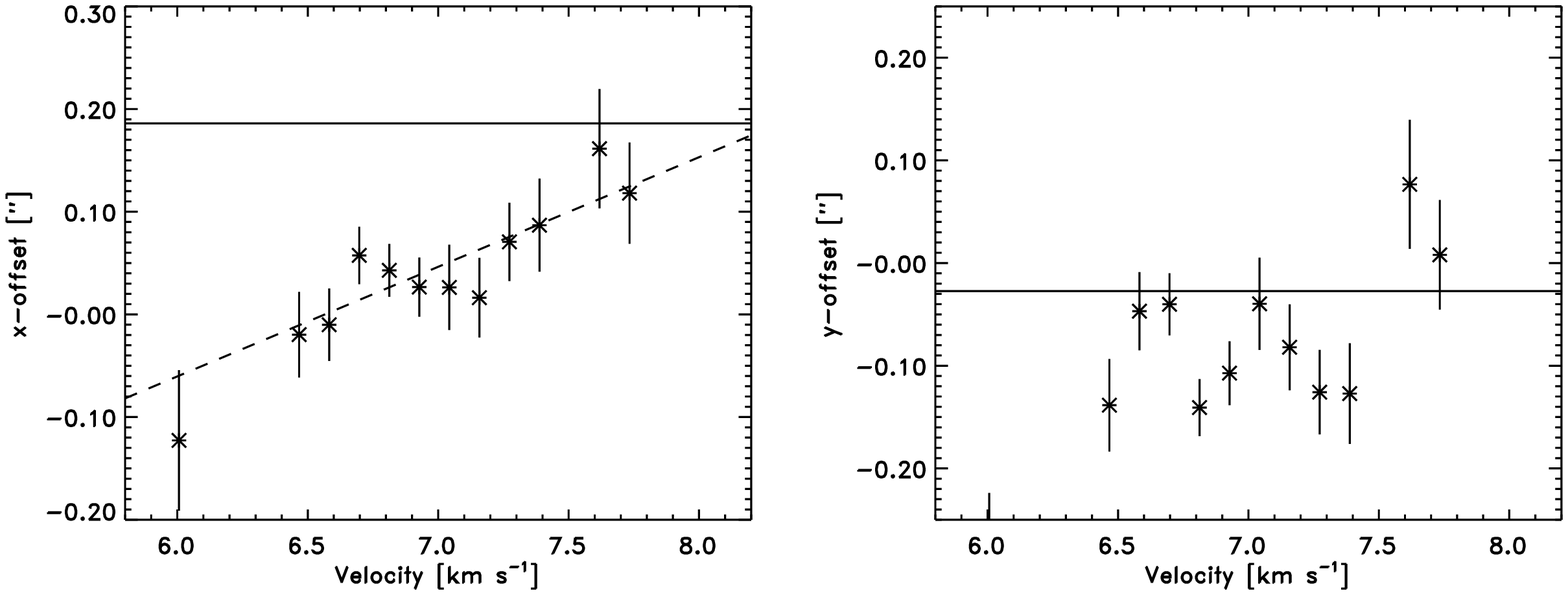}}
\caption{Velocity gradient in the H$_2^{18}$O emission toward IRAS4B:
  \emph{Upper panel:} Moment-1 (velocity) map with the direction of the
  largest gradient indicated by the solid line and the direction of
  the water masers with the dashed line and the larger-scale molecular
  outflow with the dotted line. \emph{Lower panels:} the fitted position
  offset (relative to the phase center of the observations) along
  (left) and perpendicular (right) to the direction of the main
  gradient for each channel where at least 3$\sigma$ emission is
  detected. The solid lines in both panels indicate the continuum
  position; the dashed line in the left-hand panel a linear gradient
  fit to the data.}\label{vel_gradient}
\end{figure}

The high spectral and spatial resolution offered by the IRAM data
reveals a tentative velocity gradient in the H$_2^{18}$O emission.
Fig.~\ref{vel_gradient} shows a moment-1 (velocity) map of this
emission indicating a change in velocity of a few~$\times$~0.1~\kms\
over the extent of the source emission. To estimate the magnitude of
the velocity gradient the centroid of the emission was determined
channel-by-channel in the $(u,v)$-plane. The derived offsets along and
perpendicular to the largest velocity gradient are then plotted as
function of velocity and a linear fit performed
(Fig.~\ref{vel_gradient}). The velocity gradient found in this way is
9.4~\kms~arcsec$^{-1}$ or 7.7$\times 10^3$~\kms~pc$^{-1}$ at a
3.5--4$\sigma$ confidence level. The velocity gradient is about 3
orders of magnitude larger than the typical gradients observed on
arcminute scales in pre-stellar cores \citep{goodman93} -- supporting
a scenario in which the H$_2^{18}$O emission has its origin in a more
rapidly rotating structure such as a central disk -- although it is
not possible to address whether the velocity for example is Keplerian
in nature. The position angle of the largest gradient is
$\approx$75$^\circ$ measured East of North (with an accuracy of
$\pm$5--10$^\circ$ from the gradient measured in the $(u,v)$ plane),
which, interestingly, is nearly perpendicular to the axis of the
protostellar outflow (upper panel of Fig.~\ref{vel_gradient}) traced
by water masers in the North-South direction with a position angle of
$-$29$^\circ$ \citep[][]{marvel08,desmurs09} and thermal line emission
on larger scales at a position angle of 0$^\circ$
\citep[e.g.,][]{choi01,prosacpaper}.

Besides the tentative velocity gradient, the H$_2^{18}$O line width is
remarkably narrow: its width to zero intensity is about 1.7~\kms\ with
a full width half maximum (FWHM) from a Gaussian fit of 1.0~\kms. For
comparison, a free-falling envelope toward a central star with even a
very small mass of 0.1~$M_\odot$ would have a characteristic infall
velocity of 2.5~\kms\ at 25~AU -- i.e., we would expect a line width
significantly larger than that observed in this case. A rotationally
supported edge-on disk would show a similar line width, whereas a more
face-on disk, would provide a lower width in agreement with what is
observed, although a disk seen entirely face-on would not produce any
observable velocity gradient. Alternatively, an inclined disk with
sub-keplerian rotation (e.g., a ``pseudo-disk''
\citep{galli93a,hennebelle09}) or a density enhancement in the inner
envelope due to a magnetic shock wall slowing down the infalling
material \citep{tassis05,chiang08}) could result in a lower line width
- although such phenomena usually occur on larger scales than
25~AU. \cite{watson07} argued based on the detected mid-IR emission
that the outflow cone must be close to pole-on providing a low opacity
route for the IR emission to escape toward the observer. This would
imply a nearly face-on disk. However, the proper motions and radial
velocities of H$_2$O masers suggest an outflow directed closer to the
plane of the sky \citep{marvel08,desmurs09}. Clearly observations of
the dynamical structure of the inner regions of IRAS4B with even
higher angular resolution are required to distinguish between these
scenarios.

\subsection{Column density, mass and abundance}
The strength of the H$_2^{18}$O transition suggests a significant
reservoir of water vapor toward IRAS4B. We estimate the column density
of H$_2$O by adopting an excitation temperature for p-H$_2^{18}$O of
170~K from the modeling results of \cite{watson07} and assume that the
emission is optically thin and uniform over its extent. With these
assumptions we estimate a column density for p-H$_2^{18}$O of $4\times
10^{15}$~cm$^{-2}$, which translates into a total H$_2$O column
density of $9\times 10^{18}$~cm$^{-2}$ assuming a $^{16}$O/$^{18}$O
ratio of 560 and an ortho-para ratio for H$_2$O of 3 (i.e., an
ortho-para ratio established at high temperatures). For comparison the
column density estimated based on the Spitzer detections by
\cite{watson07} is two orders of magnitude lower, 9.2$\times
10^{16}$~cm$^{-2}$, over an emitting area of
0.24\arcsec$\times$0.24\arcsec\ -- i.e., comparable to the deconvolved
size of the H$_2^{18}$O emission here. Our inferred column density is
almost unchanged if the excitation temperature is lowered to 100~K and
increases by up to a factor 5 if the temperature is increased to
1000~K. These temperatures cover the range of conditions expected for
any of the scenarios for the origin of the H$_2$O emission discussed
in \S\ref{introduction}.

The total mass contained in the detected H$_2$O is 6.0$\times
10^{-8}$~$M_\odot$, or 0.02~$M_{\rm Earth}$.  Assuming a typical
H$_2$O abundance relative to H$_2$ of 10$^{-4}$, corresponding to
sublimation of the H$_2$O rich dust ice mantles
\citep[e.g.,][]{pontoppidan04}, the total H$_2$ mass of the H$_2$O
emitting material (dust+gas) is 7.5$\times 10^{-5}$~$M_\odot$ or 25
$M_{\rm Earth}$. For comparison the mass of the compact disk around
IRAS4B inferred from the modeling of high angular resolution dust
continuum observations is 0.24~$M_\odot$ \citep{evolpaper}. Thus, if
the H$_2^{18}$O emission has its origin in this disk, it arises in a
small fraction $\approx 0.03$\% of the material in the disk.

Alternatively, in the absence of such a disk, it is possible that the
emission has its origin in the hot inner region of the protostellar
envelope where the temperature is $\gtrsim 100$~K: for a simple
power-law envelope density profile reproducing the submillimeter
continuum emission for IRAS4B on scales larger than $\sim$1000~AU
(2.8~$M_\odot$ within 8000~AU; \citealt{evolpaper}), the mass within
25~AU (where the temperature is higher than about 100~K) is about
5$\times$10$^{-4}~M_\odot$, implying a H$_2$O abundance of about
1.5$\times 10^{-5}$. However, such a model is not self-consistent on
small scales: to fit the observed compact dust continuum emission seen
by the interferometer a strong increase in the envelope density on
small scales by two orders of magnitude is required -- above the
already increasing radial density profile
\citep[e.g.,][]{evolpaper}. However, if such a density enhancement was
due to a magnetic field wall as discussed above \citep{chiang08}, the
H$_2$O abundance would be lower by the same amount, dropping to about
1.5$\times 10^{-7}$. This abundance is low compared to the expectation
from the full desorption of the H$_2$O mantles and also lower than the
constraints on the H$_2$O abundance in the outer envelopes of the
IRAS4 sources where H$_2$O is frozen-out based on ISO-LWS results
\citep{maret02}. A low H$_2$O abundance in the region of grain-mantle
desorption may reflect destruction of H$_2$O by X-rays
\citep{staeuber06}, but the H$_2$O abundance would need to be reduced
to the levels of the outer cold envelope where H$_2$O is frozen-out
and thus could not provide the compact emission observed here.

Models of the chemistry in more evolved disks around pre-main sequence
stars (where the envelope has dissipated) show a warm upper layer
where H$_2$O gas can exist. Although these models are not fully
appropriate for disks in the embedded phase, where UV photons may not
be able to freely reach the disk surface and heat the gas, they
provide a useful reference point for comparison. If the H$_2$O gas
phase abundance is just determined by the balance of photodesorption
of H$_2$O ice and freeze-out, typical gaseous H$_2$O column densities
at 10--25 AU are a few $\times 10^{18}$ cm$^{-2}$, dropping rapidly at
larger radii in these models \citep[e.g.,][]{oberg09h2o}, only
slightly lower than those found here. Alternatively, the temperatures
in the upper layers of the disk may be hot enough to drive an active
gas-phase chemistry. \citet{woitke09} find a layer of irradiated hot
water at altitudes $z/R$~=~0.1--0.3 extending out to 30 AU where
temperatures are 200--1500 K and densities $10^8-10^{10}$ cm$^{-3}$,
comparable to the conditions deduced here (see also
\cite{glassgold09}). The H$_2$O mass in this layer is $\sim 10^{-4}$
$M_{\rm Earth}$ in their model, about two orders of magnitude lower
than derived on basis of the H$_2^{18}$O observations presented in
this paper.  Because their H$_2$O/H$_2$ abundance is only
$10^{-6}-10^{-5}$, the inferred total warm H$_2$ mass is comparable.

The discrepancy between the column densities from these data and those
from the {\it Spitzer} mid-infrared observations \citep{watson07}
remains significant, though. A possible explanation is that the
mid-infrared observations are limited by extinction and thus do not
probe the total water column density. Alternatively, a lower
temperature of the H$_2$O emitting gas of $\sim 100$~K could with an
unchanged column density keep the observed H$_2^{18}$O line intensity
at the same level while decreasing the mid-IR line flux predicted by
the model.

It also remains a question why IRAS4B is the only source with strong
H$_2$O lines at mid-infrared wavelengths
\citep{watson07}. \citeauthor{watson07} proposed that IRAS4B was
either seen in a particularly favorable geometry -- with the H$_2$O
mid-infrared emission from disks around other sources being masked by
optically thick envelopes -- or that the release of H$_2$O into the
gas-phase constituted a particularly short-lived stage in the
evolution of embedded protostars. These scenarios could easily be
distinguished by further observations of H$_2^{18}$O emission from
embedded protostars at (sub)millimeter wavelengths where the envelopes
are optically thin and water should thus be detectable if present at
similar levels as in IRAS4B. Future observations with the Atacama
Large Millimeter Array in Band 5 will fully open this topic up for
ground-based observations of large samples of embedded protostars in
the same lines.

\acknowledgments We are grateful to the IRAM staff -- Arancha
Castro-Carrizo, in particular -- for help with the observations and
reduction of the resulting data, and to Michiel Hogerheijde and Ruud
Visser for information on H$_2$O gas-phase column densities in disk
models.  We are also grateful to David Neufeld and the anonymous
referee for insightful comments on the paper. The research in
astrochemistry in Leiden is supported by a Spinoza Grant from the
Netherlands Organization for Scientific Research (NWO) and a NOVA
grant.

\end{document}